# THE COLLOQUIUM ON DECOUPLING CIVIL TIMEKEEPING FROM EARTH ROTATION

John H. Seago,[*] Robert L. Seaman,[†] and Steven L. Allen[‡]

On October 5 and October 6, 2011, the Colloquium on the *Decoupling Civil Timekeeping from Earth Rotation* was hosted in Exton, Pennsylvania by Analytical Graphics, Inc. (AGI). This paper highlights various technical perspectives offered through these proceedings, including expressions of concern and various recommendations offered by colloquium participants.

**INTRODUCTION**

In response to the requests of some colloquium participants and other professionals, the co-chairs of the colloquium on *Decoupling Civil Timekeeping from Earth Rotation*[§] have written an extended introduction and review for the proceedings. This summary does not intend to provide a complete, or even balanced, presentation of the issues surrounding the topic of decoupling of civil timekeeping from Earth rotation, but primarily serves to highlight some of the more significant technical perspectives made during the meeting according to the judgment of the co-chairs. It is hoped that the highlighting of these points encourages the reader to explore these proceedings thoroughly; this summary is no substitute for reading the actual manuscripts and discussions, as there is a wealth of detailed information to be discovered by careful review.

**DECOUPLING CIVIL TIMEKEEPING FROM EARTH ROTATION**

There is an old saying that "the man with one watch always knows the time, but the man with two watches can never be sure." Implicit to this proverb is the observation that timekeepers run at different rates. For practical reasons then, some ultimate timekeeper must be declared, with all others approximating it. For example, the *leap day* on February 29th keeps the two very different astronomical cycles of *day* and *year* synchronized. It does so by varying the number of days in the calendar year to keep up with the actual orbital year of approximately 365.242 days. The need for these adjustments is appreciated by a general public having learned it from a young age, and the requirement for leap days has been known since antiquity; for without leap-day insertions, the dates of a 365-day calendar would start to lose their familiar relationship with the seasons.

Until recently, our definitive timekeeper was the rotation of the Earth, or analogously, the motion of the sky as seen from the rotating Earth. The task of determining time was therefore left to astronomers, with whom Earth rotation came to be known in the 20th century as *Universal Time* (UT), a measure which is synonymous with *mean solar time* at the meridian of Greenwich. *Mean solar time* is Earth rotation relative to the fixed stars (*sidereal time*) adjusted by approximately four extra minutes per day to keep up with the diurnal rising and setting of the Sun over the long term. The days of the calendar are thereby maintained with *mean solar days*.

---

[*] Astrodynamics Engineer, Analytical Graphics, Inc., 220 Valley Creek Blvd, Exton, PA, 19341-2380.
[†] Senior Software Systems Engineer, National Optical Astronomy Observatory, 950 N Cherry Ave, Tucson, AZ 85719.
[‡] Programmer/Analyst, UCO/Lick Observatory, ISB 1156 High Street Santa Cruz, CA 95064

[§] http://futureofutc.org/

The longstanding requirement to reconcile clocks with the mean-time rotation of the Earth has recently been questioned within communities that maintain and rely on ultra-precise frequency standards, sometimes known as "atomic clocks". Atomic resonators were invented in the 1950's, and by the 1960's ensembles of precise atomic-frequency standards were being averaged together to create laboratory time scales more uniform than the rotation of Earth. This led to a definition of duration within the metric system, known as the *Système International* (*SI) second*, in terms of a hyperfine transition frequency of cæsium. Although the *SI* second was calibrated to be close to $1/_{86400}$ of a mean solar day, the two measures of duration are fundamentally different, such that astronomical time slowly and irregularly drifts away from atomic time at an unpredictable rate. Like the man with two watches, a decision had to be made as to whether mean solar days, or atomic radiation, would be the ultimate timekeeper. Atomic frequency was important for broadcasting and telecommunications at the dawn of the space age, but astronomical time had its own technical and societal usages, with national laws and international regulations specifying that radio time-signal emissions should track Universal Time. Practically, Earth rotation cannot be adjusted; only the artificial atomic scale is adjustable—"how?" was the main question.

Atomic frequency standards enabled unprecedented synchronization of broadcasts with UT, informally known as "coordinated" Universal Time. Time had always been determined locally and astronomically; because the wireless transmission of time signals began before the International Bureau of Weights and Measures (BIPM) performed experimental work on the measurement of time or frequency, the regulation of time scales used for radio broadcasts landed with the Radiocommunication sector of the International Telecommunication Bureau (ITU-R). In 1972 regulations recommended that broadcast time signals should track the version of UT called UT1 using a new compromise which was by 1974 formally named Coordinated Universal Time (UTC).[*] Under this recommendation, a sequence of atomic seconds known as International Atomic Time (TAI) was continually maintained in the background by averaging the readings of many laboratory frequency standards. However, once Earth time (UT1) and atomic time (UTC) drifted about one-half second apart, the last day of a month was necessarily lengthened by one second by international agreement. This adjustment became known as the *leap second*, which serves the same purpose in clock time as a leap day serves calendar time.[1] Its scheduling is announced by the International Earth Rotation and Reference System Service (IERS) months in advance, and the notice is distributed electronically throughout telecommunications systems using services such as satellite navigation systems and via the internet using Network Time Protocol (NTP).

A proposal will come before the ITU-R Radiocommunication Assembly in January 2012 on whether to decouple civil timekeeping from Earth rotation by ceasing leap seconds from UTC after 2017. This proposal would allow UTC to diverge from UT1 without bound. The points of debate surrounding this proposal are quite numerous and now well-rehearsed from all sides.[2] The primary function of this colloquium, however, was to discuss the ramifications of such a proposal should it succeed, because the adverse impacts from redefining UTC have not been extensively researched. The colloquium chairmen were delighted with the breadth and quality of the contribu-tions and the enthusiasm of the contributors, particularly through the thoughtful round-table discussions. It is therefore all the more remarkable that the general topic of civil timekeeping, and the specific issue of whether it should remain coupled to Earth rotation, have received so little airing outside of a few very narrow technical communities.

---

[*] Universal Time (UT) defines time-of-day on Earth. There are several realizations of UT, including UTC. UT1 best represents the instantaneous orientation of Earth about its rotational axis.

**WHAT IS CIVIL TIMEKEEPING?**

It would be misleading to now claim that civil timekeeping is based on the seasonally non-uniform time-of-day indicated by the length of shadows, also known as *apparent solar time*. Modern civilization moved from absolute reliance on sundials centuries ago to favor artificial timepieces that keep uniform time. Within the context of "decoupling civil timekeeping from Earth rotation," the technicalities of the meaning and measurement of Earth rotation as Universal Time (UT) were understood and appreciated within the technically inclined audience of the colloquium. Yet the more general topic of *civil timekeeping* is harder to compartmentalize.

In its broadest sense, *civil timekeeping* implies "time kept for civilization." However, this is much more than a standard to which the man-on-the-street sets his wristwatch. Timing signals are used by electronic devices which drive and control modern infrastructure, such as communication, transportation, finance, energy, and defense systems. Civil timekeeping involves both calendar-date stamping and high-precision frequency generation. The technological usage of civil time constrains its specification.

As civil timekeeping is considered within the service of today's culture, the most susceptible activity may be the programming of computing devices that drive the technology of modern society. This activity now has decades of legacy code deeply embedded, some of which operates as seemingly inaccessible "black boxes". There is a need to recognize that this is the way the world operates today, and that significant technology is built around the existing civil-timekeeping standard—both correctly and incorrectly. Although difficult, overcoming the "software inertia" of black boxes is possible, but it requires dedicated resources.

At the same time, there is also a need to recognize that computers work for people; people don't work for computers. Clocks have been invented and maintained by humanity for very specific reasons. Civil time cannot be defined arbitrarily and still be indefinitely useful; the rate of clocks must approximate mean solar time if date stamps from clocks are to remain useful for civil timekeeping purposes over the long term. The existing paradigm appears to operate well within our increasingly network-connected society, which raises doubts as to whether the continued coupling of civil timekeeping and Earth rotation presents significant or imminent problems.

**THE CONCERNS OVER DECOUPLING**

The topic of the colloquium relates to a very fundamental change, as this appears to be the first time in human history that civil timekeeping would be decoupled from Earth rotation *by design*. Historically, timekeeping has been principally perceived as an exercise in remaining faithful to what the sky was doing. Although timekeeping practices may have occasionally gone awry throughout recorded history, there was always willingness to make a correction once the errors became noticeable.

*Are we creating unknown problems for the **present** with the current proposal to decouple civil timekeeping from Earth rotation*? Based on limited declarations of ITU-R study groups, it is unclear how thoroughly certain technology domains have been represented and assessed by groups studying the issue of UTC redefinition. The technical discussions have been framed in terms of the elimination of leap seconds, which limits participation to those stakeholders who know about leap seconds, understand the difference between UT1 and UTC, and appreciate the implications of that relationship. The low number of responses to surveys, data calls, and other targeted inquiries suggests that the discussions have been too technical for most non-experts to appreciate. Many people outside the IERS especially lack awareness of this issue, with most astronomers not likely understanding the difference between UT1 and UTC. Unfortunately, the cessation of leap seconds would create complications for some organizations, notably those related to military defense, as well as astrodynamics and astronomy.

*Are we creating unknown problems for the **future** with the current proposal to decouple civil timekeeping from Earth rotation*? History suggests that astronomical timekeeping has always been specially regarded and there is no compelling evidence that it will be disregarded in the future. Rather, pressure may be eventually brought about from an innate principle of "astronomical conformity" which will lead to a desire for societies to recouple civil timekeeping and Earth rotation sometime in the future. How can we help future generations prepare for that recoupling now? Alternative schemes involving very infrequent intercalary adjustments (such as a leap minute or leap hour) cannot be credibly presumed to work once short-term adjustments are formally abolished. As technology becomes increasingly complex, it would seem extremely difficult to resynchronize civil time with Earth rotation if they were allowed to separate, and that the level of difficulty would grow with magnitude of the difference.

When our descendants ask why a fundamental change to human timekeeping was made in the early 21$^{st}$ century, the reasoning that was followed should be obvious, credible, and justifiable. Yet the study process in place is opaque and the motivations for UTC redefinition are not absolutely clear. For example, the ITU-R authorized a Study Question which decided that any recommendation for change should be based on a determination of user requirements, but that determination was never concluded. Some issues might never be acknowledged publicly; nevertheless, studies by the government agencies have generated tomes of public documentation for much less consequential questions. One would therefore expect voluminous documentation exploring the motivations, impacts, and repercussions across affected communities that one might normally expect out of a decision-making process. However, that documentation apparently does not exist, at least in the public domain, and therefore the information on which national governments are basing their decisions is unclear.

Because the existing process within the ITU-R has not clearly determined end-user requirements, organizations outside this process have attempted to poll for this information. Over the past decade, published survey results have expressed high favor for the *status quo* and rather low regard for the alternative of ceasing leap seconds. But survey attempts stimulate controversy as a means to collect opinions, inform people, and create awareness, and survey results have been largely dismissed by those who feel that such polling provides no useful information. Also, most surveys have been conducted "informally" to avoid the difficulty of gathering more official positions, so the opinions of decision-makers who must also consider the financial aspects may not be reflected. It is not clear that official decision-making processes within governmental organizations have been sufficiently transparent or that the most affected communities have registered substantive input.

## THE CONSEQUENCES OF DECOUPLING

### Impacts on Hardware

Celestial tracking and data acquisition are usually accomplished via multiple systems in which timekeeping signals are exchanged via complex messaging protocols. Telemetry is logged and maintained over many years, and provided to diverse users for different technical purposes which are often coordinated with other tracking systems. The relevant usable time intervals for such data may range from fractions of a second to decades (or longer). Tracking systems assuming UTC ≈ UT1 would need to be re-implemented to explicitly distinguish between UTC and Earth rotation (UT1), or isolated to receive a vetted UT1 input, or be retired or replaced. Systems already accommodating a UT1-UTC correction would have to be vetted for proper operation given values of |UT1-UTC| > 0.9s, possibly from a new source. Even systems unaffected by a decoupling of UTC and UT1 would have to be inventoried and assessed (with significant expense) to prove this.

**Impacts on Software**

Only a minority of software properly accounts for leap seconds, including software for real-time systems; nevertheless, it is still an outstanding question as to how much and to whom this matters. It is possible that the affected codebases of systems assuming UTC ≈ UT1 are much smaller than those that do not make such an assumption, and it is also possible that there are substantial costs for maintaining leap seconds. Yet, each of these possibilities argues for software inventories that have yet to happen. Some preliminary inventories (*e.g.*, for astronomical software systems) forecast major complications should UTC no longer closely approximate UT1.

There is agreement that timekeeping software must be more flexible than in the past, but what this means depends on one's viewpoint regarding UTC redefinition. For those who favor the decoupling from Earth rotation, "flexible software" should not assume that UT1-UTC is bounded in the future. For those respecting the current system, "flexible software" maintains the necessary infrastructure to accommodate small intercalary adjustments past, present, and future.

*Would the decoupling of civil timekeeping from Earth rotation result in the development of more or less "flexible" software for the future*? The cessation of future leap seconds does not mitigate the need to account for past leap seconds in software, so any argument that conventions must change to continue with badly written software is unpersuasive. Convenience to software programming is a weak reason for denying our progeny civil time linked to the astronomical day; programming mistakes will be made regardless of whether civil time is coupled or not to Earth rotation, so the future of civil timekeeping should not hinge on the anticipated incompetence of some programmers now or in the future.

**Example Application: Impacts on GNSS**

A convincing case can be made that existing global navigation satellite system (GNSS) programs would not be benefited, but would be financially impaired, by UTC redefinition. These systems already function through leap seconds and redefinition of UTC would require changes to operational software, established procedures, documentation, and testing. Such systems are relatively inflexible due to the criticality of their missions, such that seemingly inconsequential changes can be costly. Such costs would be unavoidable and borne by the taxpaying public.

**Technical Confusion**

Polling suggests that people are apprehensive about changing something that has worked for a very long time, and the one thing that is generally understood by people with any sensibility of timekeeping and time scales is that UTC is approximately UT1. A change in the meaning of time is therefore guaranteed to cause confusion, even in situations where systems do not malfunction. Service providers cannot control how users interpret standards or whether they are aware of any changes to implicit operating assumptions.

The proposed decoupling poses a special challenge to providing future data in software and almanacs that are a function of the rotational angle of the Earth. If data continue to be provided as a function of Universal Time, an intensive re-education strategy will be required. Users would thereafter need to clearly understand the difference between UTC and UT1 and know how to obtain and apply UT1-UTC from external sources. For example, even if the impact of dropping leap seconds were manageable for practicing celestial navigators through re-training and future almanac updates, a greater risk arises from the potential for confusion in literature, textbooks, and other educational materials already published for a subject that is a rarely exercised backup.

The creation of accurate technical documentation is a significant expense that is often neglected or deferred. The necessary documentation changes would disturb technology domains that are otherwise unaffected by a redefinition of UTC; therefore, the documentation aspects are likely to be very far reaching and extremely tricky to manage. A redefinition of a fundamental scientific time scale cannot avoid leaving permanent signatures in time-series taken before and after the change; this also would be a source of continuing confusion into the future. The technical definition of an atomic time scale called "UTC" first with, and then without, leap seconds will be complicated. Civil standards that do not approximate Universal Time would best avoid the label "Coordinated Universal Time" and its acronym "UTC", because these descriptions have always implied a technically useful realization of Universal Time.

**Societal Confusion**

The astronomical basis of timekeeping fulfills its social function even when it is not perfectly observed. This is because general perceptions are what count within the realm of symbolism, not just the facts. Many people would perceive a decision to decouple civil timekeeping from Earth rotation as an exercise of authoritarian power that no elite scientific or technical group should have been able to influence. Even if the effects of such a change were not immediately apparent or very gradual for most purposes, individuals would still be upset and would talk about the decoupling from Earth rotation as if timekeeping were already broken.

**Educational Confusion**

There is remarkable public fascination with the notion of leap seconds which can be leveraged to promote public awareness and emphasize the relevance of higher education in science, technology, engineering, and mathematics. Explanations of leap seconds often involve the Moon, tides, and gravity, making the topic a rewarding introduction into the domain of solar-system dynamics and its direct relevance to people on Earth. Leap seconds are an effect that an ordinary person can observe themselves, because a precise watch will reveal the one-second difference after a leap second compared to a GNSS receiver or a radio-controlled clock. Similarly, sundials are wonderful objects to teach the seasons or the inequality of the days, Kepler's laws, or the practical application of trigonometry. Therefore, from a pedagogical point of view, the decoupling of civil time from that kept by the Sun can be seen as an additional complexity that would not be easy to explain to pupils and the general public.

**Legal Confusion**

Some national governments recognize the astronomical basis of timekeeping explicitly, while others recognize it implicitly. The existing system is the only one known to be reconcilable with all existing national and international statutes, and a civil timekeeping standard decoupled from Earth rotation has no known precedent. Detailed consideration of various legal implications appears to be lacking; for example, future disruptions caused by adjustments larger than one second could spawn many legal complications.

**DISCUSSED ALTERNATIVES**

Alternatives to maintaining leap seconds have been proposed over the years, but options that keep civil time coupled to Earth rotation have been largely dismissed, offering no improvement over the existing functional system. Various options have included: relaxing the tolerance for |UT1-UTC|, redefining the *SI* second, and inserting required adjustments at less frequent but regular intervals.[3] Variations on these alternatives were discussed at the colloquium, although these variations do not represent all possible options.

**Predictions from a Quadratic Model**

A low-order (quadratic) model of the separation of Earth rotation (UT1) from uniform time could be used to make predictable adjustments to the calendar decades or centuries in advance. These corrections, based on a computational rule rather than empirical observation, would not necessarily be accurate enough to satisfy technical applications requiring UT1 from clocks, but such corrections would facilitate long-term software development and maintain a general perception of astronomical coupling. However, several concerns were noted with this approach.

1. Decadal variations can be significantly larger than the long term quadratic trend, making a quadratic model too inaccurate for the intended purpose in the near term.

2. Although extended predictions of UT1-UTC might help some software applications, software issues are primarily caused by complete ignorance of intercalary adjustments, rather than a need for their predictable forecast.

3. Having a predictable leaping rule for time of day still does not necessarily solve the software issues. (For example, programmers sometimes code the Gregorian calendar rules incorrectly.)

4. Loosening the existing ±0.9-second tolerance specification for UT1-UTC will still result in operational issues.

Regardless, such a proposed approach begs for continued geophysical studies to make Earth-orientation predictions more reliably accurate.

**Use TAI Internally and Expose *Status-Quo* UTC Externally**

It is clear that many astrodynamics, astronomical, navigational, and telecommunications applications already use this option operationally, or a variation of it (*e.g.*, GPS time in lieu of TAI seconds). It seems reasonable to promote a conventional sequence of TAI seconds without leap seconds as an internal scale, and as a source of determining precision time interval whenever necessary, as all modern scales are already functionally related to, or approximate the rate of, TAI. Prior to 2001, the internal use of TAI was recommended by the Director of the BIPM, the Consultative Committee on Time and Frequency (CCTF), and by the ITU-R through multiple Recommendations. The use of TAI-like time has been mainly hindered by the lack of transmission of DTAI = TAI–UTC, but the availability of GNSS scales through navigation signals may be overcoming this limitation. A technically interesting approach to implementing this method involves the synchronization of a computer's operating kernel to a uniform background scale and tracking the difference between UTC and the background scale (*e.g.*, TAI or GPS) using the widely deployed Time Zone Database.[*]

**Increasing the Intercalary Interval**

The idea of making the necessary adjustments at much longer intervals, for instance at the end of each century, is appealing in that most people do not live that long and therefore would have to

---

[*] http://www.iana.org/time-zones

deal with the problem either once or never. However, the following drawbacks of this proposal were noted:

1. New software is being written all the time so it is not simply a once-per-century matter. Software developers of the future are just as likely to focus on more immediate problems and may not expect their software to still be in use a few decades into the future, although such software might very well become entrenched as other code is built around it.

2. There is a real risk that any long-term adjustment scheme will not be reliably implemented when the declared time comes because long-term proposals such as a leap minute, leap hour, centennial adjustments, *etc*. push the technicalities sufficiently far into the future such that the recoupling would not be pragmatically addressed.

3. Future large adjustments still have the problem of labeling events during the adjustment interval in an atypical way, just as with leap seconds, except their introduction will be harder to ignore.

4. A change from the current standard will still result in immediate operational issues for some.

Perhaps the most surprising notion mentioned was whether the current proposal implied that humanity was supposed to wait until |UT1-UTC| approximated one day, at which point an unscheduled leap day would be introduced into the calendar. Such a scenario admits that there are really only two types of calendar adjustments presently, leap seconds and leap days, and abandonment of the smaller adjustment leaves humanity with only unrealistic and impractical options employing the larger, which is insufficient to address fundamental issues. For example, the current proposal to decouple creates a potential civil issue where a Saturday could turn into a Sunday; however, the seven-day week is a universally accepted convention based on mutually agreed tradition and is important to millions of faithful across the globe as well as the general populace.

**Other Options Not Discussed**

Feedback regarding adverse impacts from redefining Coordinated Universal Time was a primary goal of the meeting, which is a prerequisite to entertaining alternate options. However, it is not the only prerequisite, such that additional discussions, public meetings, and engineering-planning activities surrounding these issues should be encouraged for the future.

## RECOMMENDATIONS FROM COLLOQUIUM DISCUSSANTS

No attempt was made to formalize any consensual recommendation(s) of the attending group because there was no specific professional, technical, or governmental entity requesting such recommendations from this gathering. Nevertheless, several suggestions were advanced by various colloquium attendees, some of which are now highlighted.

**Reconsideration of the Issue**

Action by the ITU-R at this time would seem unwise based on the potential widespread impacts of their decision and the lack of documentation supporting the need for a change. Therefore, a delay in any official decision is advisable. Consideration of civil timekeeping issues or adoption of UTC stewardship by other international entities or standards organization should be explored. The coherent collection of requirements and application of systems engineering best practices should provide a framework for reaching consensus.

### Contact with National Governments

Colloquium participants recognized that a decision to change the *status quo* would impact various communities outside telecommunications. Responsible person(s) within government departments should be made aware that there is much more at stake than ramifications and consequences to telecommunications.

### Transmission of TAI-UTC

DTAI (TAI-UTC) should be made more readily available. Transmission of DTAI would allow it to be received autonomously, avoiding haphazard changes by hand which is burdensome for systems where configurations are tightly controlled. Prior leap seconds are part of the historical record and whether or not UTC is redefined, a database of intercalary adjustments will be perpetually required for some applications. Access to the latest entry of this database is equivalent to "transmission of DTAI", and this requirement now persists however UTC is defined in the future.

### Distributing the Offset between Civil Time and Earth Rotation

If Coordinated Universal Time is redefined to no longer be coordinated with Universal Time, the user base interested in UT1-UTC corrections would likely increase suddenly. For this reason, there must be some means of ensuring practical distribution of UT1-UTC, especially for machine usage. Suitable standards-based technologies exist but these must have a long expected lifetime, be practically implemented by both the producer and the consumer, and maintain system integrity and data security. An XML-based service is one possible approach.

Methods of distribution should start out as simple as possible, such as file access via http, and care must be exercised to avoid proposed solutions that cause unnecessary complications. For example, the idea of introducing an NTP service of variable frequency that attempts to track UT1 (as UTC did before 1972) might create unnecessary complications. Generally, it seemed that transmission of UT1-UTC corrections would be preferable. The IERS Directing Board should be involved in this issue, although the IERS is primarily a confederation of separately funded geophysical research centers rather than a transmission center for timing products. The recommendation to make UT1-UTC more readily available stands independently of whether civil timekeeping is decoupled from Earth rotation, because many operations need a realization of Earth rotation that is more precise than provided by civil clocks.

### Extended Prediction of Leap Seconds and UT1-UTC

If leap seconds were retained, it seems that the IERS may already be able to confidently predict leap seconds two or three years in advance. Increasingly advanced prediction of UT1-UTC would be a good option for almanacs and other publications that are physically printed, regardless of whether civil timekeeping is decoupled from Earth rotation.

### Development of Specialized Analysis Tools

If Earth rotation and civil time are decoupled, the development of specialized simulation tools would be necessary to provide some sense of the adverse operational impacts or to simulate the magnitude of symptoms caused by decoupling. There also seems to be a need for specialized diagnostic tools to assist in discovering where UTC is used as a realization of Earth rotation within complex operational systems and software.

### Engagement of Additional Stakeholder Communities

Civil timekeeping is an issue of interest to all and efforts should be made to broadly engage possible stakeholder communities. If UTC is redefined, additional meetings should be organized for the future, some of which perhaps having narrower focus on more topical issues (astronomy, astrodynamics, *etc*.) to address the technical challenges that will result from the decoupling of civil timekeeping from Earth rotation.

**CONCLUSION**

While numerous points of view were well expressed by the contributors of this colloquium, a few points stood out in the opinion of the chairmen. Primarily, the motivations for decoupling civil timekeeping from Earth rotation are not entirely apparent, and the supposed advantages to making a change have not been shown to outweigh the supposed disadvantages both now and in the future. A coarser coupling between civil timekeeping and Earth rotation would result in adjustments of less frequency and larger magnitude, which would be more noticeable and less practical to maintain than the current system having predictable insertion points (at end of the month), constrained values (-1, 0, or 1), and a prescription for tagging events occurring during the adjustment (23:59:60). Because the existing UTC system with leap seconds is already implemented, it should be strongly preferred over alternative proposals or protocols that lack obvious advantages.

The historical tendency of timekeeping practices has been to move from empirically observed adjustments to predictable ones based on calculation. Perhaps the creation of a predictably accurate time-of-day adjustment algorithm, valid for at least a decade or more, might provide a viable alternative, but this has not been seriously studied up to the present. It appears that Earth rotation may now be predictably accurate to within one second out to three years with 99% confidence. If so, perhaps we should continue to seek advances that would improve the prediction of Earth rotation with high confidence out to one decade or better, with the goal of eventually replacing empirically predicted leap seconds with an algorithmic correction to time-of-day. Such an approach acknowledges that the coupling of civil timekeeping and Earth rotation is fundamentally an issue of calendar maintenance, and thereby satisfies public expectations of preserving astronomical relationships. Yet, because this approach may not be able to provide a sufficiently accurate realization of Earth rotation from clocks, it seems best to leave the current system in place until such time that Earth rotation models are improved, and for systems that prefer a uniform internal time scale to use IEEE 1588-2008 (Precision Time Protocol) with GNSS time or other time scale closely tied to TAI.

Finally, an outstanding question is whether "UTC without leap seconds" really represents a permanent scenario for decoupling. A persuasive argument can be made that the expectation of civil-timekeeping's astronomical basis is so deeply ingrained in our society that a departure from it is unlikely to last forever. If so, an attempted decoupling now may be the worst of all possible options, because it creates issues for both present users reliant upon the existing system and it causes unforeseeable complications for generations to come. The want of the current decoupling proposal to address the most fundamental issues erodes its favorability within many technical communities and the public, resulting in a lack of consensus.

**REFERENCES**

[1] Finkleman, D., S. Allen, J.H. Seago, R. Seaman and P.K. Seidelmann (2011), "The Future of Time: UTC and the Leap Second." *American Scientist*, Vol. 99, No. 4, p. 316.

[2] Finkleman, D., J.H. Seago, P.K. Seidelmann (2010), "The Debate over UTC and Leap Seconds". Paper AIAA 2010-8391, from the Proceedings of the AIAA/AAS Astrodynamics Specialist Conference, Toronto, Canada, August 2-5, 2010.

[3] McCarthy, D.D., W.J. Klepczynski (1999), "GPS and Leap Seconds—Time to Change?" *GPS World*, November, pp. 50–57.